\documentclass[CEJP,PDF]{cej} 

\usepackage{amssymb}
\usepackage{graphicx}
\usepackage{subfigure}
\usepackage{textcomp}
\usepackage[tight]{units}
\usepackage{layout}

\title{Structural Changes in Data Communication in Wireless Sensor Networks}

\articletype{Editorial}

\author{Raquel S. Cabral \inst{1} \email{raquelcabral@gmail.com},
        Andre L. L. Aquino \inst{2} \email{alla.lins@gmail.com},
        Alejandro C. Frery \inst{2} \email{acfrery@gmail.com}\\
        Osvaldo A. Rosso\inst{2}$^,$\inst{3} \email{oarosso@gmail.com},
        Jaime A. Ram\'irez \inst{1} \email{jramirez@ufmg.br}
        }

\institute{
     \inst{1} Graduate Program in Electrical Engineering - Federal University of Minas Gerais - Av. Ant\^onio Carlos 6627, 31270-901, Belo Horizonte, MG, Brazil\\ 
     \inst{2} Scientific Computing and Numerical Analysis Laboratory - Federal University of Alagoas - Av. Lourival Melo Mota, s/n, 57072-900, Macei\'o , AL, Brazil \\
     \inst{3} Complex Systems Laboratory, Engineering Faculty - University of Buenos Aires - Paseo Col\'on 850,  Buenos Aires, Argentina \\
      }

\abstract{
Wireless sensor networks are an important technology for making distributed autonomous measures in hostile or inaccessible environments.
Among the challenges they pose, the way data travel among them is a relevant issue since their structure is quite dynamic.
The operational topology of such devices can often be described by complex networks.
In this work, we assess the variation of measures commonly employed in the complex networks literature applied to wireless sensor networks.
Four data communication strategies were considered: geometric, random, small-world, and scale-free models, along with the shortest path length measure.
The sensitivity of this measure was analyzed with respect to the following perturbations: insertion and removal of nodes in the geometric strategy; and insertion, removal and rewiring of links in the other models. 
The assessment was performed using the normalized Kullback-Leibler divergence and Hellinger distance quantifiers, both deriving from the Information Theory framework. 
The results reveal that the shortest path length is sensitive to perturbations.
}


\begin{document}
\maketitle

\section{Introduction}
\label{sec:introduction}

Wireless Sensor Networks (WSNs) are an emerging technology that allows the monitoring of physical variables, such as temperature, sound, light, vibration, pressure or movement~\cite{Akyildiz2002}. 
A WSN consists of a large number of wireless autonomous devices, called ``sensor nodes'', ``sensors'' or ``nodes.'' 
These entities work in a cooperative way sensing the environment, communicating among them, and taking decisions. 
Such networks are a promising technology in a wide range of applications, for instance, biotechnology, industry, public health, and transportation~\cite{Arampatzis2005}.
They are of particular interest for monitoring hostile or inaccessible environments~\cite{OverviewWirelessSensorNetworksTechnologyEvolution}.
	
\begin{figure}[h]
\centering
    \subfigure[Flooding based]{\includegraphics[width=.45\textwidth]{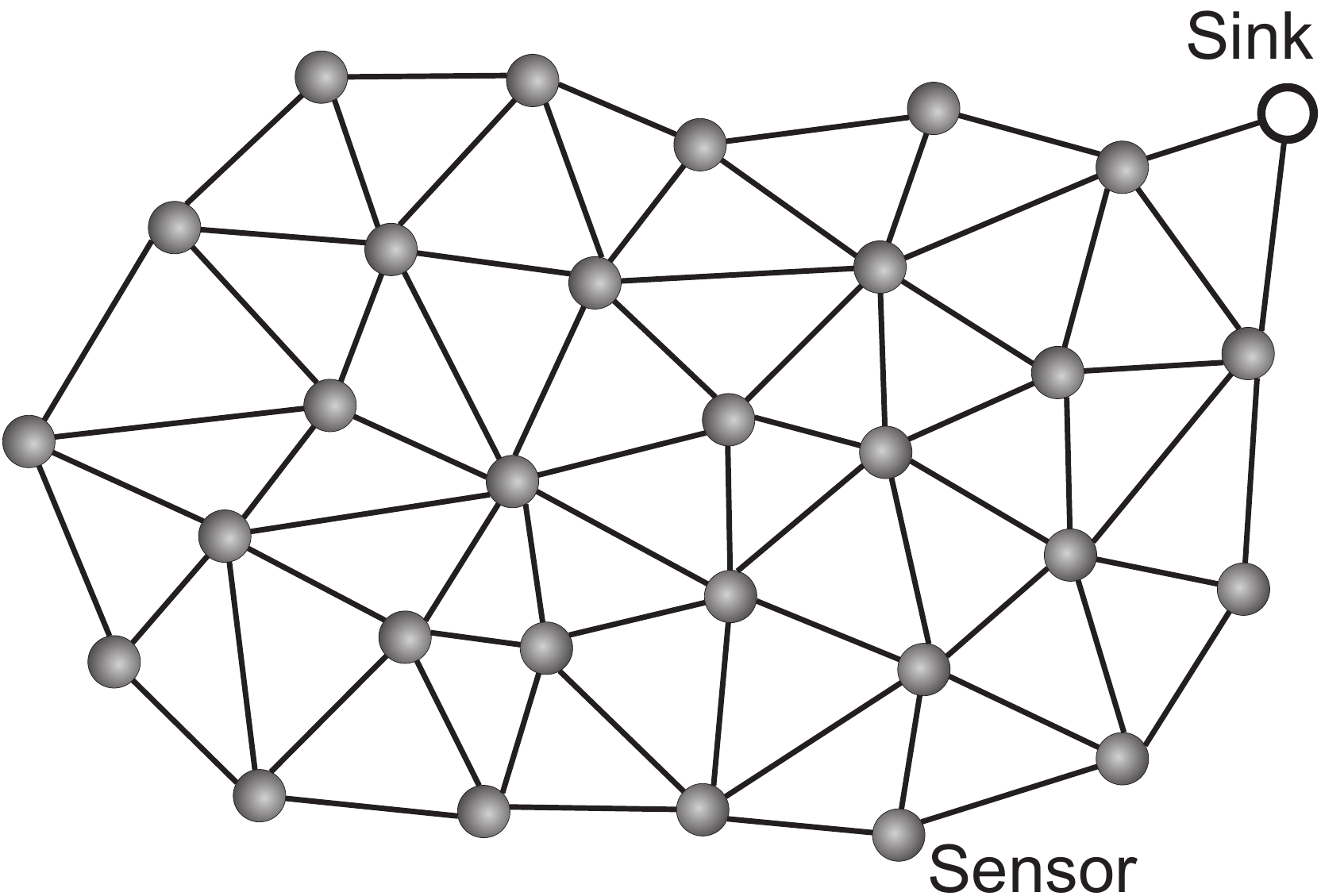}
    \label{fig:estrutura_enlace}}
    \hspace{4mm}
    \subfigure[Complex network based with scale-free model]{\includegraphics[width=.45\textwidth]{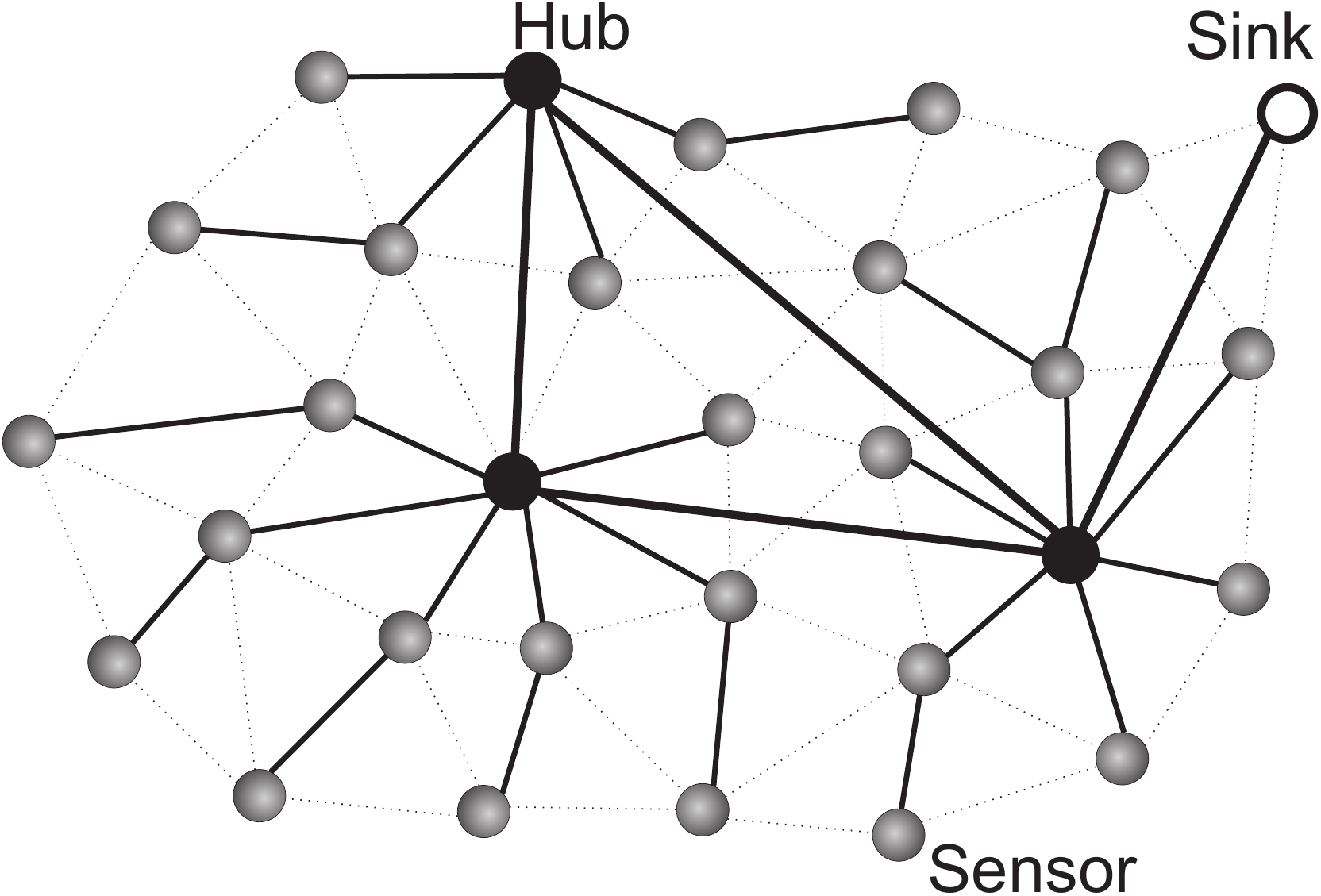}
    \label{fig:estrutura_overlay}
    }
    \caption{Examples of WSNs communications}
    \label{fig:topologies}
\end{figure}

The physical variables are monitored and stored in the sensor, and propagated to a sink node. 
The sink is a management node responsible for processing the data and delivering it to an external user~\cite{DataDrivenPerformanceEvaluationWirelessSensorNetworks}. 
The communication uses the nodes between the sensor source and sink in an ad-hoc fashion. 
The two most common communication strategies are based on data flooding and on complex networks. 
In the \textit{flooding based}, the data communication starts from a node to its direct neighbors, then each neighbor re-propagates to the next neighbors, and so on. 
Each node propagates each information once and this processes repeats until the data arrives to the sink node.
Fig.~\ref{fig:estrutura_enlace} illustrates this communication strategy. 

In \textit{complex network based}, the communication depends on the model used to characterize it (random, small-world or scale-free, for example).
So, specific topological properties of complex networks are used to determine this kind of communication. 
Fig~\ref{fig:estrutura_overlay} shows a complex network based on scale-free model.
An important issue in this model is the presence of nodes with high degree known as ``hubs''. 
The communication is towards the nodes closer to the sink or to a hub. 
The hubs can propagate the information to distant nodes and, consequently, the information flow is concentrated on them.

The WSN communication strategies can be characterized by a set of measures.
These measures describe different features on the network, such as connectivity, centrality, cycles or distances~\cite{Costa2008}. 
Example of measures are the average shortest path length, clustering coefficient, network diameter and betweenness. 
In this work, we use the shortest path length, because it is directly related to the network energy consumption and sudden changes in this measure may result in an increase or decrease in energy consumption.

Energy availability is a critical feature in WSN~\cite{Akyildiz2002}.
Moreover, classical routing solutions in WSNs employ the shortest paths~\cite{Akkaya2005}. If the shortest paths change, the routing solutions may become inefficient.

The communication structure in WSNs is dynamic, i.e., it changes over time.
It is, therefore, important to analyze how this reflects on the measures.
The most common changes are: addition and inactivity of nodes, in the flooding based communication; and insertion, removal and rewiring of links, in the complex network based communication~\cite{Boas2010}.
These changes have a great impact over real WSNs design. 
The infrastructure mechanisms usually are designed considering topological characteristics. 
For instance, in a routing algorithm based on shortest path length if a node is removed the routing behaviour is affected.

In this way, the main question of this work is: 
\begin{quote}
\textit{``What is the impact of changes or perturbations on these strategies, as measured by the shortest path length?''}
\end{quote}

This work presents the analysis of communication strategies in WSNs by means of analyzing the variation of measures.
We analyze flooding, random, small-world, and scale-free networks, and the shortest path length. 
The variation of measures was analyzed with respect to the insertion and removal of nodes in flooding; and with respect to insertion, removal and rewiring of links in the strategy based in complex networks. 
Stochastic quantifiers, namely the normalized Kullback-Leibler divergence and Hellinger distance~\cite{Nascimento2010}, were used to quantify the variation of the measure. 
The results reveal to which extent the measure is influenced by the perturbations considered: the shortest path length exhibits a clear dependence on the type and intensity of the perturbation.

The paper is organized as follow.
Section~\ref{sec:related_work} presents the related work. 
Section~\ref{sec:wsnTopologies} analyses communication strategies in WSNs.
Section~\ref{sec:results} presents the results of measures behavior and the quantifiers performance.
Finally, conclusions and future directions are presented in Section~\ref{sec:conclusion}.

\section{Related work} 
\label{sec:related_work}

This section presents a brief review about WSNs communication strategies that use complex networks concepts, and stochastic quantifiers applied to networks.

\subsection{WSNs communication strategies}

Helmy~\cite{Helmy2003} described WSNs as spatial graphs that tend to be much more clustered and with higher path lengths than random graphs. 
He showed that it is possible to reduce the path length of wireless networks with the addition of a few short cut links. 

Guidoni et al.~\cite{Guidoni2010} proposed on-line models to design heterogeneous sensor networks with small world features. 
The proposed model takes into account the data communication flow in this kind of networks to create shortcuts towards the sink in such a way that the communication between the sink and the sensor nodes is optimized. 
The network presents better small world features when the shortcuts are created, and an interesting trade off between energy and communication latency is observed. 

Ruela et al.~\cite{Ruela2010} improved the data communication by using hubs, an approach based on scale-free networks.
The hubs introduce useful characteristics of complex networks, e.g. small average shortest path length between all sensors and the sink, and high clustering coefficient.
This strategy saves resources, avoiding excessive communication and, consequently, reducing the time to data delivery.

\subsection{Information Theory quantifiers}

Wang et al.~\cite{Wang2006} used entropy to provide an average measure of networks heterogeneity since it measures the diversity of the link distribution.

Boas et al.~\cite{Boas2010} analyzed the effect of perturbations in complex networks. 
They choosed measures based on the fact that the network characterization is made from samples rather than from the entire network.
They applied three perturbations: link addition, link removal, and link rewiring. 	

Carpi et al.~\cite{Carpi2011} proposed a new quantifier based on Information Theory for the analysis of dynamic network evolution.
It is used to compute changes in topological randomness for degree distribution of the network. 
The quantifier, a statistical complexity measure, is  obtained by the product of the normalized Shannon entropy and the normalized Jensen-Shannon distance.

In this paper, we use two stochastic quantifiers, a divergence and a distance, to quantify the changes in WSNs.
We show that it is possible to identify the strength of the perturbations that leads to a breakdown of the network properties, i.e, a kind of phase transition.
Moreover, we evaluate the flooding communication strategy considering node perturbations. 
These aspects are not addressed in the related works mentioned and constitute the main contributions. 

\section{Analysis of communication strategies in WSNs}
\label{sec:wsnTopologies}

The WSN physical connectivity can be described as an undirected graph $\mathcal G = (\mathcal{V},\mathcal{E})$, where $\mathcal{V} = \{v_1, v_2, \ldots, v_N\}$ is the set of sensor nodes, and $\mathcal{E} = \{(v_i, v_j): 1 \leq i, j \leq N\}$ is the set of links between nodes. 
Each link is determined by the communication geometry. 
In our case it is a circumference, i.e., the neighborhood of node $v_i$ is formed by all nodes $v_j$ at most $R$ units away from $v_i$.
The data employ this physical connectivity to propagate, but obeying the rules of a protocol.
The protocol may impose restrictions on the use of the available links $\mathcal E$, resulting in a communication subgraph $G=(\mathcal G, E)$, such that $E\subset \mathcal E$ and, therefore, $G\subset \mathcal G$.

The number of connections of a node is its degree $\kappa$.
An important graph characteristic is its degree distribution, i.e., the collection of all its degrees.

The \textit{flooding based communication} employs all the available links, so $G=\mathcal G$.
The flooding starts propagating the data from $v_i$ to all its neighbors, then each neighbor re-propagates to its neighbors, and so on. 
Each node propagates each unit of information once. 
This processes repeats until the data arrives to the sink node.

The communication based on \textit{complex networks} employs a subset of links $E\subset \mathcal E$ such that data flow towards the nodes which are closest to the sink.
The source nodes always send the information to that neighbor which reduces the distance to the sink. 
Each node propagates the information following the same criterion.
This process repeats until the data arrives to the sink node. 

In some cases the path to the sink is not the optimal, because it is not always immediate to find the optimal routing solution in distributed scenarios.
These scenarios offer only local information, a limitation which often prevents finding global optima. 
Specific measures of complex networks are used to characterize this kind of communication, for instance, the shortest path length~\cite{Newman1999,Boccaletti2006}. 
A shortest path is any path that connects two nodes and has minimal length.
It is an important measure in communication strategies. 

There are several complex networks models. 
In this work, we use the random, small-world, and scale-free models:
\begin{description}
\item[Random model] The probability to connect each pair of nodes is the same. 
There are two ways to build a random graph~\cite{Erdos1960}: $N$ nodes and exactly $M$ links, and $N$ nodes and the probability $0<p_c<1$ to connect each pair of nodes.
In the first description, $M$ links are uniformly distributed among the $N(N-1)/2$ possibilities.
In the second description, which is the one we adopted here, we start with a totally disconnected graph and then connect each pair of nodes with probability $p_c$. In this case, the probability of observing $0\leq k\leq N-1$ connections in each node follows a Binomial distribution with $N-1$ trials and probability of success $p_c$, leading to $(N-1)p_c$ as mean degree.

\item[Small-world model] A communication strategy is small-world when the communication network presents a high clustering coefficient and a small shortest path length~\cite{Watts1998}. 
There are several ways to build such strategy; in this work, we use the Watts-Strogatz model. 
This model starts with a circular regular topology with $N$ nodes, each one connected to the $k$ nearest neighbors in each direction (right or left in circular topology). 
Then, each link is randomly ``rewired'' with probability $p_{r}$~\cite{Costa2008}, i.e., if the current link is $(v_i,v_j)$ and there is no link $(v_i,v_k)$, rewiring consists in deleting $(v_i,v_j)$ and creating $(v_i,v_k)$.

\item[Scale-free model] A communication strategy is scale-free when the communication network displays a power law degree distribution $p(\kappa) \sim \kappa^{-\lambda}$, $\kappa>0$, with $2 < \lambda < 3$. 
The main feature of this topology is the presence of some nodes with high degree, often called ``hubs''. 
To generate this topology we use the Barab\'{a}si-Albert scale-free model that starts with a small number of nodes with $m_0$ links, and in the next step a new node with $m$ links is added to the network (with $m\leq  m_0$).
The probability connection is linearly proportional to the node degree~\cite{Barabasi1999}.
\end{description}

The main question stated in this work is: 
\textit{``What is the impact of changes or perturbations on these strategies, as measured by the shortest path length?''}
To answer this question, we need to quantify the variation of the measure when perturbations occur.
The most common changes, or perturbations, are: the addition and inactivity of nodes, in the flooding based communication; and insertion, removal and rewiring of links, in the complex network based communication. 

The stochastic nature of the aforementioned quantifiers suggests the use of techniques deriving from Information Theory in order to assess their change.
The normalized Kullback-Leibler divergence and Hellinger distance are two quantifiers suitable for describing the difference between distributions~\cite{Nascimento2010}.

Consider the discrete random variables $X$ and $Y$ defined on the same sample space $\Omega = \{\xi_1, \xi_2, \dots, \xi_n \}$.
The distributions are characterized by their probability functions $p, q \colon \Omega \to [0,1]$, where $p(\xi_i) = \Pr(X=\xi_i)$ and $q(\xi_i) = \Pr(Y=\xi_i)$.
A metric $\mathcal D$ between these two distributions is a quantifier obeying:  
\begin{enumerate}
\item\label{prop:reflex} $\mathcal{D}(p,p) = 0$, reflexivity; 
\item\label{prop:nonnegat} $\mathcal{D}(p,q) \geqslant 0$, non-negativity; 
\item\label{prop:commute} $\mathcal{D}(p,q) = \mathcal{D}(q,p)$, commutativity; 
\item\label{prop:triangle} $\mathcal{D}(p,q) \leqslant \mathcal{D}(p,r) + \mathcal{D}(r,q)$, triangle inequality for any other probability function $r$ defined on the same probability space.
\end{enumerate}
A distance is not required to satisfy property~\ref{prop:triangle}, and a divergence is only required to satisfy properties~\ref{prop:reflex} and~\ref{prop:nonnegat}~\cite{Cha2002}.

Assuming $q(\xi)>0$ for every event $\xi\in\Omega$, the Kullback-Leibler divergence is defined as
\begin{equation}
 D_{KL}(p,q) = \sum_{\xi\in\Omega} p(\xi) \log \frac{p(\xi)}{q(\xi)} .\label{eq:entropiaRelativa}
\end{equation}

The Hellinger distance does not impose positivity on the probabilities; it is defined as
\begin{equation}
 \mathcal{D}_H(p,q) = \frac{1}{\sqrt{2}} \sqrt{
 \sum_{\xi \in \Omega}\Big( 
 \sqrt{p(\xi)} - \sqrt{q(\xi)}
 \Big)^2
 } = 1 - \sum_{\xi\in\Omega}\sqrt{ p(\xi) q(\xi)}.
\label{eq:hellinger}
\end{equation}

In order to make the Kullback-Leibler divergence (an unbounded positive quantity) and the Hellinger distance (which is confined to the $[0,1]$ interval) comparable, in the remainder of this work we will use the normalized Kullback-Leibler distance defined as ${\mathcal D}_{KL}(p,q)=1-\exp\{-D_{KL}(p,q)\}$.

\section{Simulation results} 
\label{sec:results}

This section presents the simulation study about the complex network measures behavior of WSNs using two quantifiers presented previously.

\subsection{Methodology}

Simulation assumptions and parameters were:

\begin{description}

\item[Network parameters] In order to simulate a sparse WSNs, we used $N = 1000$ nodes deployed in an $L^2 = 100\times100$ area.
The communication radius of each node was $R = 5$ units in strategies based in flooding communication. 
With these values, we got a network density approximated $1.5$ obtained through $d = \pi\, R\, N/L^2$, where $d$ represents the number of neighbours of each nodes~\cite{Alla_2012}.
The probability of connection in the random model was $p_c = 0.06$, which implies an average degree equal to $6$.
The nearest neighbors and probability rewiring in the small-world model were $k = 3$ and $p_r = 0.3$, respectively.
The number of links added in each step in the scale-free model was $m = 1$.
The above parameters were chosen according to Boas et.al. \citep{Boas2010}.

\item[Perturbations] 
In the  strategies based in flooding communication we performed: (i)~\textit{nodes addition}, nodes were randomly added to the network; and (ii)~\textit{nodes removal}, nodes were randomly removed from the network, that represents the node inactivity.
The perturbations were performed on $\{1\%, 2\%,  \ldots, 10\% \}$ of the total number of nodes.
In the complex networks based communication, we performed tree types of links perturbations~\cite{Boas2010}: (i)~\textit{link removal}, links were randomly removed from the network; (ii)~\textit{link addition}, two unconnected nodes were randomly selected, and a new link was established; and (iii)~\textit{link rewiring}. 
In this last case, the perturbations were performed in $\{1\%, 2\%,  \ldots, 10\% \}$ of the total number of links.

\item[Normalization] for each network, the normalized histogram $\mathcal{H}$, also known as histogram of proportions, was obtained with 200 bins of equal width. 
The $\mathcal{D}_{KL}$ diverges for $q(\xi) = 0$ and $p(\xi) \neq 0$, as defined in equation~\eqref{eq:entropiaRelativa}. 
In order to avoid the division by zero, a small positive constant $\delta = 0.001$ was added
to each bin, and then the histogram is normalized to add $1$~\cite{Boas2010,Cabella2008}. 
The original histogram is used to compute the Hellinger distance once it does not impose the positivity restriction on the probabilities.

\item[General parameters] for each communication strategy, we generated 10 different networks and for each network 100 different perturbations were made. 
In this way, we are able to present the mean results with symmetrical asymptotic confidence intervals at the $95\%$ significance level.

\item[Computational resources] we performed the evaluation using the \texttt R platform \cite{Rmanual}, on an Intel(R) Core(TM) i5 CPU 760 \unit[2.80]{GHz}  with \unit[7]{GB} RAM, running Ubuntu 12.04 (\unit[64]{bits}).
The \texttt{igraph} library was used to generate and modify the graphs~\cite{igraph}.
\end{description}

Figures~\ref{fig:lattice_D_GRG_l} and~\ref{fig:lattice_D_ER_WS_BA_l} present the variation of the shortest path length, for the different communication strategies and the perturbations considered.
Each plot presents two quantifiers: the normalized Kullback-Leibler divergence $\mathcal{D}_{KL}$ (denoted as ``$\triangle$''), and the Hellinger distance $\mathcal{D}_{H}$ (denoted as ``$\circ$''), as functions of the level (intensity) of the perturbation. 

\subsection{Flooding communication}

\begin{figure}[hbt]
\begin{center}
\includegraphics[width=1\textwidth]{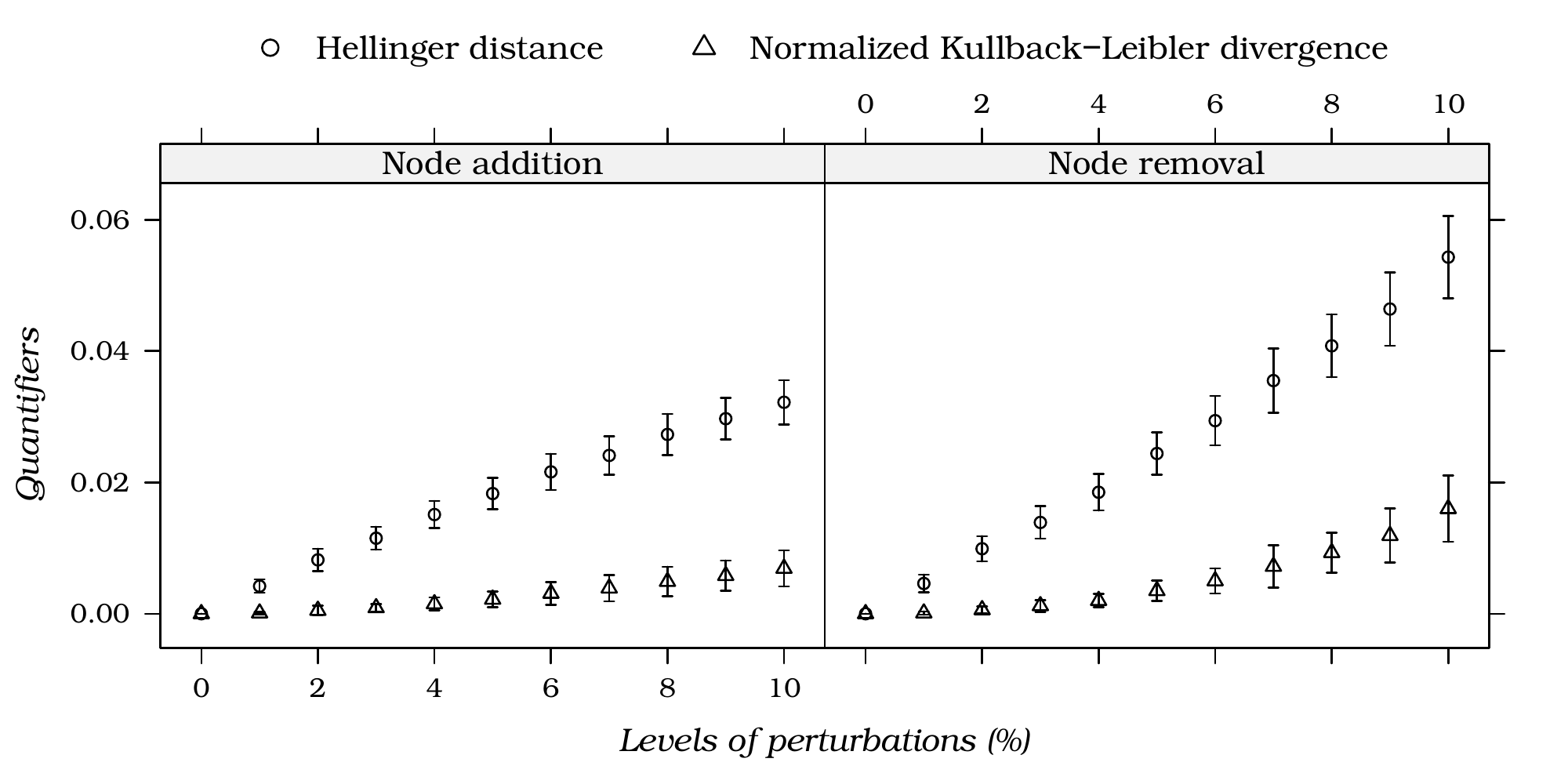}
\caption{Hellinger distance $\mathcal{D}_{H}$ and Kullback-Leibler divergence $\mathcal{D}_{KL}$ for the shortest path length to the geometric network model (GRG).}
\label{fig:lattice_D_GRG_l}
\end{center}
\end{figure}

Figure~\ref{fig:lattice_D_GRG_l} shows the shortest path variation in flooding communication; from left to right, node addition and node removal.
Both addition and removal alter the shortest path length, the bigger the level of perturbation the stronger the change.
The Hellinger distance is more sensitive to changes in the shortest path length than the normalized Kullback-Leibler divergence.
The variability of both measures has some dependence on the level of perturbation; it increases with the level of perturbation but then stabilizes.
The removal of nodes has stronger impact on the shortest path length than node addition (the former variation is steeper than the latter); this effect is more noticeable with higher levels of perturbation.
This may be explained by the fact that when new nodes are added within the communication range of a node, the shortest path is little affected.
The shortest path length changes when nodes are removed because many links are lost and some of them belonged to the shortest paths.

The quantifiers are, thus, sensitive to both node addition and removal in this scenario, but with different responses.
The Hellinger distance is the one which varies most, reaching, in mean, $0.032$ for the addition and $0.054$ for the most intense removal of nodes ($10\%$).

The results presented are important to WSNs designers. Based on this study, a more efficient topology control can be applied. For instance, when 10\% of nodes are added the shortest path length changes. This change indicates that adding more than 10\% of nodes in the original network could be harmful. When 5\% of nodes die (i.e. are removed) the shortest path length changes. This information can be used by the designer to calibrate the duty cycling operation, where each node periodically switches between sleeping mode and awake mode~\cite{SlBr_2010}.

\subsection{Complex network communication}

\begin{figure}[hbt]
\begin{center}
\includegraphics[width=1\textwidth]{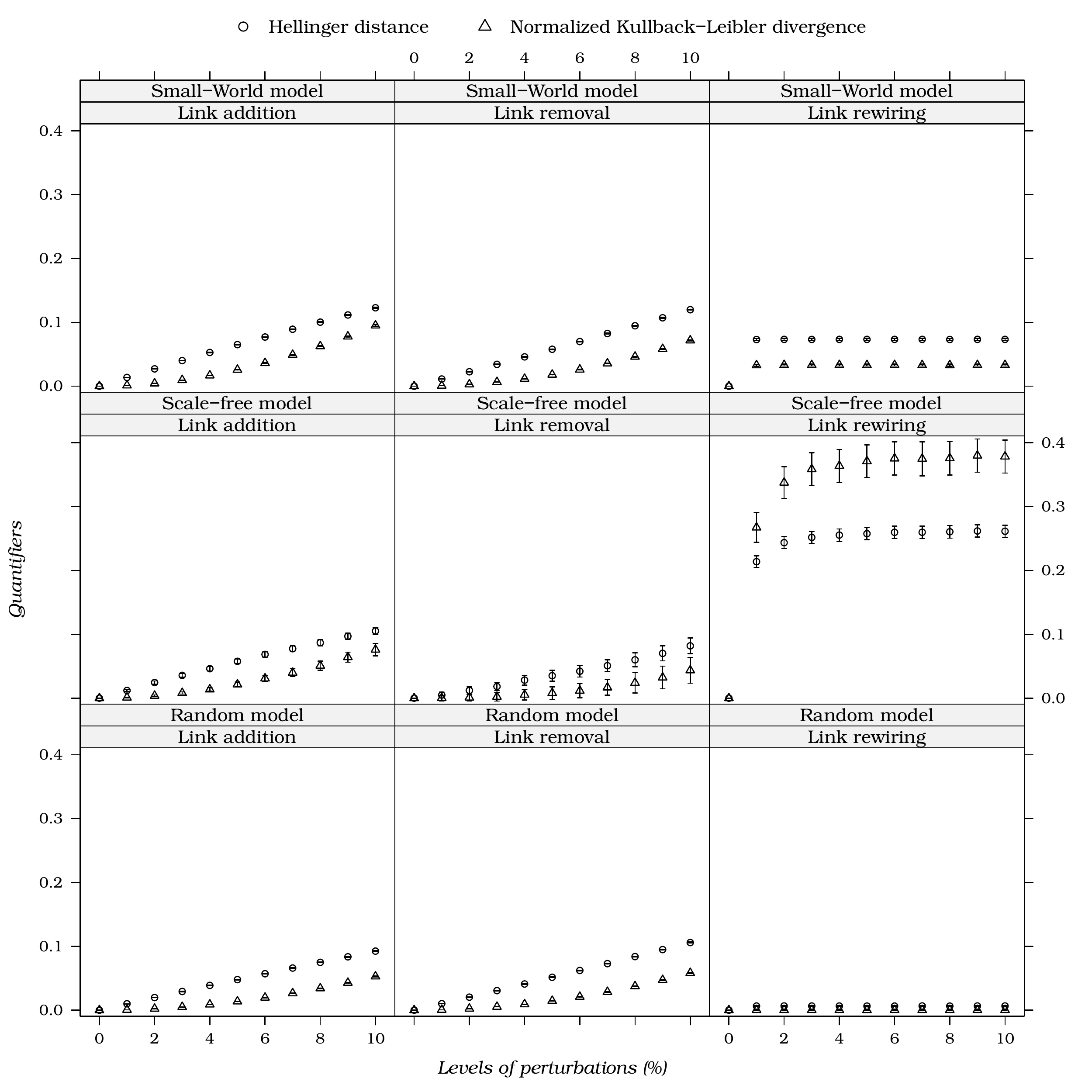}
\caption{Hellinger distance and normalized Kullback-Leibler divergence for the shortest path length to the random network model, small-world network model and scale-free network model.}
\label{fig:lattice_D_ER_WS_BA_l}
\end{center}
\end{figure}

Figure~\ref{fig:lattice_D_ER_WS_BA_l} presents the results in the random, small-world and scale-free communication models (top to bottom rows) and the three types of perturbations: link addition, link removal and link rewiring (left to right columns).
The shortest path length is sensitive to the link addition and removal applied in the three communication models, and both quantifiers ($\mathcal{D}_{KL}$ and $\mathcal{D}_{H}$) behave alike: the stronger the perturbation, the more the quantifier changes in direct proportion.

Regarding these two perturbations, again, the Hellinger distance exhibits more intense variations with respect to the perturbation than the normalized Kullback-Leibler divergence; notice that the confidence intervals do not overlap.
Table~\ref{tab:mean_values_quantifiers} shows the mean value of the most intense perturbation ($10\%$); this corresponds to the rightmost point of the plots of Figure~\ref{fig:lattice_D_ER_WS_BA_l}. 
The Hellinger distance is consistently and significantly higher than the Kullback-Leibler divergence by a factor of, approximately, two, in all the situations where link addition and removal are applied.

The behavior of link addition and removal is the same in the tree models.
Both affect the shortest path length, since these perturbations increase or decrease the distance between nodes. 
In particular, networks where all nodes have the same degree (regular graphs) may become small-world with just a few reconnections~\cite{Watts1998}, i.e., randomly adding or removing links may result in connecting nodes which are far away, reducing the shortest path length.
Link rewiring has a different behavior.
Although the quantifiers exhibit differences, see Table~\ref{tab:mean_values_quantifiers}, they are negligible in the random model.
Link rewiring alters small-world models, but its effect seems constant, i.e., independent of the intensity of the perturbation.
Regarding the scale-free model, there is a strong variation of both measures when small perturbations are applied, but the change tends to stabilize, i.e., saturates, soon after.
The former is the only case where the normalized Kullback-Leibler divergence is bigger than the Hellinger distance.

The main feature of these networks is the presence of hubs, which makes them sensitive to this kind of perturbation: the shortest path length alters radically whenever a link involving hubs is added or removed.
Additionally, the removal of some links makes the network disconnected. 

\begin{table}[hbt]
\caption{Mean most intense perturbation ($10\%$) in the Hellinger distance and normalized Kullback-Leibler divergence.}
\label{tab:mean_values_quantifiers}	
\centering
\begin{tabular}{@{}c|*3{r}|*3{r}@{}}
\hline
 & \multicolumn{3}{c|}{Hellinger distance} & \multicolumn{3}{c}{Normalized Kullback-Leibler divergence} \\  
 & Link addition & Link removal & Link rewiring & Link addition & Link removal & Link rewiring \\ \hline
Small-world model & 0.122 & 0.119 & 0.073 & 0.095 & 0.071 & 0.033 \\ 
Scale-free model & 0.105 & 0.082 & 0.261 & 0.076 & 0.043 & 0.378 \\ 
Random Model & 0.092 & 0.105 & 0.006 & 0.052 & 0.058 & $10^{-4}$ \\
\hline
\end{tabular}
\end{table}

All models analysed represent a routing topology generated by some management strategies. Generally, these strategies combine the shortest path length with other QoS parameter (energy, delay or priority). The results reveal that when 5\% of the links are added, or removed, the shortest path length changes. This information could be used to calibrate the management strategies to avoid the interference of QoS parameter during the routing generation, once this task considers just the links addition or removal.

\section{Conclusion} \label{sec:conclusion}

The analysis of the variability of measures in WSNs communication strategies provides important information.
It gives an insight of the behavior of the network when it is perturbed and it helps us in the design of appropriate solutions for every application.

In this paper, we used the normalized Kullback-Leibler divergence and the Hellinger distance to compare three communication models: flooding, random, small-world and scale-free (the last three are members of complex network based strategies). 
We performed two types of perturbations in the flooding-based strategy: addition and removal of nodes. 
In the complex network strategy, we used three types of perturbation: addition, removal and rewiring of links. 
We analyzed how the shortest path length changes with respect to different levels of each perturbation. 

The shortest path length is sensitive to these changes, and in most situations it alters accordingly to the intensity of the perturbation.
The analysis allows identifying the relationship between the strength of the perturbations and the change of the shortest path length.
The use of quantifiers that involve logarithms or ratios may not be a good choice for this kind of characterization, because the occurrence of zeros leads to numerical problems and, possibly, to incorrect interpretation of network changes.

In addition, more efficient topology control or routing strategies in WSNs can be proposed. For instance the duty cycling or routing operations can be based on shortest path length sensibility. In the specifics scenarios treated, these mechanism could be calibrated when 5\% -- 10\% of the nodes or links are added or removed. The impact that the observed results have on the design and operation of WSNs was commented.

\section*{Acknowledgement}

This work is partially supported by the Brazilian National Council for Scientific and Technological Development (CNPq) and the Research Foundation of the State of Alagoas (FAPEAL).
O. A. Rosso gratefully acknowledges support from CONICET, Argentina and CNPq fellowship, Brazil.

\bibliographystyle{cpc}

\end{document}